\documentstyle[11pt]{l-aa}
\begin{document}
\title{ A Possible Model for the 
 Supernova/Gamma-Ray Burst Connection }

\author{ X.Y.~Wang\inst{1}, Z.G.~Dai\inst{1,2}, T.~Lu\inst{1,2},
D.M.~Wei\inst{3,4} and Y.F.~Huang\inst{1,2} }

\offprints{ T.~Lu (E-mail: tlu@nju.edu.cn) }

\institute{ Department of Astronomy, Nanjing University, Nanjing 210093, 
	    P.R. China
\and        LCRHEA, IHEP, CAS, Beijing, China
\and   Purple Mountain Observatory, Nanjing, 210008, China
\and  National Astronomical Observatories, Chinese Academy of Sciences, China
	  }

\thesaurus{ 13.07.1; 02.19.1; 08.14.1; 08.19.4} 

\date{Received  ; accepted }

\maketitle
\markboth{}{}

\begin{abstract}
Conversion from neutron stars to strange stars as a possible mechanism of
cosmological gamma-ray bursts (GRBs) has been discussed in previous works,
although the existence of strange stars is still an open question.
On the basis of this mechanism, we here outline an explanation
of the connection between supernovae (SNe) and GRBs, which has got increasing
evidence recently. 
An asymmetric but normal SN explosion leaves a massive 
($\geq1.8{\rm M_\odot}$) and rapidly rotating neutron star,
which then converts to a strange star few days later, due to its rapid 
spindown. The accompanied fireball, which can be 
accelerated  to ultra-relativistic
velocity ($\Gamma_0\sim 100$) due to the very low baryon contamination
of the strange star, flows out along the direction 
of the high-velocity SN jet and
subsequently produces a GRB and the following low energy afterglows
by interacting with the surrounding stellar wind.
We will also expect a very luminous supernova like SN1998bw,
if a  large fraction of the conversion energy  finally turns into the 
kinetic energy of the supernova
ejecta.


\end{abstract}
\section { Introduction}

More than a year ago, Galama et al. (1998) reported the detection of a very
luminous Type Ic supernova (SN) SN1998bw in the error box of GRB980425.
The estimated chance probability of the coincidence is $10^{-4}$, suggesting
a connection between these two events. From the radio observations of
GRB980425/SN1998bw, Kulkarni et al. (1998) concluded that there exists
a relativistic shock (bulk Lorentz factor 
$\gamma\equiv(1-{\beta}^2)^{-1/2}\geq{2}$ )
even 4 days after the supernova explosion. Li $\&$ Chevalier (1999)
modelled the radio light curves and inferred that the late rise observed
at days 20-40 is the result of energy input from a central engine.
These both strengthen the link between SN1998bw and GRB980425.
More recently, Bloom et al. (1999) and Reichart (1999) revisited GRB980326
and GRB970228, respectively, and found the evidence for a supernova
in the light curve and late spectral energy distribution of the afterglow.
Galama et al. (1999) reached the same conclusion for GRB970228 by the
reanalysis
of its optical and near-infrared afterglow. It appears that some
GRBs, if not all,  are connected with SNe. Now the proposed models include
``collapsar'' (Woosley 1993; MacFadyen $\&$ Woosley 1998; 
Woosley et al. 1999) or ``hypernova'' ({\rm Paczy${\acute n}$ski} 1998)
 models, in which the GRBs and SNe are powered
by the accretion process into the black hole. Some authors (e.g. Wang \&
Wheeler 1998;
Cen 1998; Nakamura 1998;)
also proposed that
the MHD prosess in the core of an asymmetric SN may be responsible for the
connction, based on the idea that the core-collapse process
is intrinsically strongly asymmetric.

Conversion from neutron stars to strange stars as a 
possible source of cosmological
GRBs has been suggested by some authors (e.g. Cheng \& Dai 1996; Ma \& Xie
1996).
Witten (1984) speculated some time ago that  strange quark matter may be 
the true ground state for hadrons (i.e. the energy of strange matter is lower
than that of matter composed of nucleus).   Farhi \& Jaffe (1984) computed
the zero temperature thermodynamics of  strange matter and found that
it may indeed be stable if the parameters of the MIT bag model take values
inside a wide ``stability window" they found. If this hypothesis were true,
the
conversion from neutron stars to strange stars may liberate a large amount
of energy ($\sim 10^{52}{\rm erg}$).
Strange stars, composed of this kind of quark matter,
have been used to explain 
some astronomical phenomena (e.g. Cheng et al. 1998; Cheng $\&$ Dai 1998;
Xu et al. 1999),
although some arguments against the 
existence should also be kept in mind (e.g. Caldwell \& Friedman 1991;
Kluz${\acute n}$iak 1994).
The conversion of a neutron star to a strange star may require the formation
of a strange matter seed, which is produced through the deconfinement of 
neutron matter at a density (Baym 1991) of $\sim7-9\rho_0$ (where $\rho_0$
is the nuclear
matter density), much larger than the central density of a $1.4M_\odot$
neutron star with a moderately stiff to stiff equation of state(EOS).
(For a soft EOS, the deconfinement density is lower; but we here assume that
the EOSs in neutron stars are moderately stiff or stiff, as requried by 
some astrophysical processes; for details please see Cheng \& Dai (1996)).
Cheng \& Dai (1996) proposed that some neutron stars in low-mass X-ray
binaries can accrete sufficient mass ($\geq0.4{\rm M_\odot}$) and undergo
 a phase transition to become strange stars. In this paper, we suggest that
a normal SN explosion may leave a massive and rapidly rotating neutron star,
which can convert to a strange star in quite a short 
time ($\sim 1 {\rm days}$) due to its rapid spindown and subsequently
produce a GRB.
We will place emphasis on examining the connection between SNe and GRBs.
If the GRBs can be sucessfully produced  few days after the SN explosion, then
 the SN-GRB connection can be reasonably explained.

\section { Ultra-relativistic jet from the strange star as GRB}

Let us consider a massive progenitor with mass greater than $20M_\odot$ on
the main
sequence that undergoes 
a Type Ib/c or Type II supernova explosion and leaves a rapidly
rotating neutron star with mass about $1.8-2.1M_\odot${\footnote{ This mass
has included that of the accretion matter from the possible supernova
fallback, see the
section Discussions.}}. 
(But, for very massive
progenitors with masses larger than $30M_\odot$,  the collapsing
iron cores may possibly implode to black holes  rather than neutron 
stars before the explosions develop,
which is just the ``collapsar" model of GRBs.)
Static neutron star with such a large mass may have
undergone phase transition to become a strange star, but for a very rapidly
rotating one (close to the break-up
angular speed), its central density may be much lower than the deconfinement
density.
Indeed, for moderately stiff EOSs, rapid rotation can sustain an extra mass
up to
$0.3M_\odot$ for a given central density (Cook et al. 1994). 
However, due to the rapid loss of angular
momentum through magnetic dipole radiation, the newborn, massive neutron
star spins 
down and its central density become larger and larger and finally 
may reach the deconfinement density  and converts to a
strange star.
The computation of Cook et al. (1994), in which they numerically study the 
rapidly rotating neutron stars in the frame of general relativity and based
on the realistic modern EOSs, has given quantative evidence for reaching
such a high density in the core of  massive neutron stars as the angular
velocities decrease. Moreover, these findings hold for a wide range EOSs,
from moderate to stiff.
To illustrate more clearly,
we here give an example from  their computation . For 
the modern EOS named FPS (Lorenz et al. 1993),
the maximum angular velocity of a $1.8629M_\odot$ (gravitational mass)
neutron star is $\omega=0.88749
\times10^{4} {\rm s}^{-1}$. The corresponding central
density is $\rho_{c}=1.4835\times10^{15}$ 
g~cm$^{-3}$. When it spins down to 
a slow motion, its central density is close to $\rho_c=3.3900\times10^{15}$
g cm$^{-3}$, having reached the deconfinement density of neutron  matter.

 Modelling of the optical light curve of SN1998bw shows that 
the time of core collapse coincides with that of GRB980425 to 
within a few  days (Iwamoto et al. 1998).
This means that the newborn, rapidly rotating neutron star 
should spin down in this timescale. If we adopt the usual magnetic dipole
radiation with the
spin-down timescale 
$T={\omega}/{\dot{\omega}}\sim2 {\rm~ days}~(\omega/10^4~s^{-1})^{-2}
(B/3.9\times10^{13}{\rm G})
^{-2}$ for typical neutron star values
(where $I$, $B$ and $R$ are, respectively, the moment of inertia, magnetic
field strength
and stellar radius of the neutron star), a nascent magnetic field larger than
$B\geq3.9\times10^{13}$ Gauss is needed. If the initial angular velocity of
the newborn neutron star is very close to the maximum (Kepler) value, 
$\Omega_K \simeq \frac{2}{3}\sqrt{\pi{G\bar{\rho}}}\simeq 8185
s^{-1}(M/{1.5M_{\odot}})^{1/2}
(R/{10{\rm Km}})^{-3/2}$, its rotational kinetic energy $E_{rot} \sim
\frac{1}{2}I\omega^2$
can be as large as $8 \times 10^{52}$erg for a neutron star with $I\sim
2\times10^{45}$
g~cm$^{2}$. This energy, if released in the form of electromagnetic waves,
 will be absorbed by the supernova ejecta (Pacini 1967; Dai \& Lu 1998) and
increase the
kinetic energy of the ejecta above what has been claimed for the SN 1998bw
(Iwamoto et al. 1998).
Fortunately, recent studies (e.g. Andersson  1998; Lindblom et al. 1998) show
that the emission of gravitational radiation due to r-mode instabilities in
hot,
young neutron stars can spin down the stars efficiently. Their calculations
show that
such stars can spin down to a few per cent of the Kepler-limit within one
year.
Recently, Ho \& Lai (1999) show that the spin-down 
may depend on the combined effect of both the r-mode instability and 
the magnetic dipole radiation, if the nascent magnetic field is sufficiently
high. The exact time when the phase transition  occurs is difficult to be
determined, 
but in
principla, it can be as short as a few days. Thus, the large initial
rotational energy 
is supposed to be lost mainly in the form of gravitational waves rather
than the usual 
electromagnetic waves. 

Once the deconfinement density is reached,
strange matter seeds are formed in the interiors of the star
and the strange matter will begin to swallow
the neutron matter in the surroundings. 
While it has been proposed that the combustion corresponds to the 
slow mode (Olinto 1987),
later work (Horvath \& Benvenuto 1988) shows that this mode is  
hydrodynamically unstable. Thus the conversion
should more likely proceed in a detonation mode (Lugones et al. 1994) (at a
speed of sound) and the timescale for the conversion is about
0.1 millisecond. 
Although neutron star is composed of outer crust, inner crust and core,
only the
outer crust will not convert into strange matter because it does not
contain free
neutrons.
Then the resulting strange star has a thin crust with 
mass $M_0\sim2\times10^{-5}M_\odot$ 
(strictly say, this mass corresponds to a $1.4M_\odot$ strange star; for a
more massive
strange star, it may increase slightly; Glendenning $\&$ Weber 1994; Huang
\& Lu 1997).
It has been pointed out (Cheng $\&$ Dai 1996) that the energy deposition of
this phase transition
is mainly through the process of $n+\nu_e\rightarrow{p+e^-}$ and $p+{\bar
\nu_e}\rightarrow
{n+e^+}$ and the phase transition energy released ($E_0$) is of the order
of $10^{52}$
ergs. The process, $\gamma\gamma\leftrightarrow{e^{+}e^{-}}$, will
inevitably lead to 
the creation of a fireball, which expands outward, carrying the baryonic
matter in the thin crust of the strange star. Finally, an
ultra-relativistic shell
is formed with a high Lorentz factor
\begin{equation}
\Gamma_0 \sim \frac{E_0}{M_{0}{c^2}} \sim 300({E_0}/{10^{52} {\rm
ergs}})({M_0}/{2\times
10^{-5}M_{\odot}})^{-1}.
\end{equation}
In addition, there's also another important process that could sometimes 
act as the central engine of GRBs
after the birth of the strange stars: differential rotation may occur in
the interiors of these newborn strange stars due to the fact that the density
profile of a strange star is much different from that of a neutron star with
the same mass (Glendening 1997). 
According to the basic idea of the Klu${\acute z}$niak $\&$ Ruderman (1998),
Dai $\&$ Lu (1998) argued that such differentially rotating strange stars
could lead to a series of subbursts
of GRBs by the following mechanism: as one part of the star rotates around
the 
other part,
internal poloidal magnetic field will be wound up into a toroidal
configuration
and linearly amplified . 
When the  toroidal field increases up to a critical field value,  it will
be able to float up to break through
the stellar surface. Reconnection of the newborn surface magnetic field will
arise a quickly explosive event with a large amount energy, which could
lead to a subburst of a GRB.
The obvious advantage in these two scenarios is that
the baryon contamination 
in the fireball is small
 due to the low mass of the crust of  strange stars.

Can this ultra-relativistic shell(s) produce a detectable $\gamma$-ray
burst inside the
dense SN ejecta of supernova? It depends on the
scattering optical opacity of ejecta. The scattering optical depth
is $\tau=\sigma_{T}nl\sim 9.4(M_{ej}/M_{\odot})(t/10{\rm
days})^{-2}(v/{3\times10^{9}{\rm cm~s^{-1}}})^{-2}$.
For Type II supernova with a ejecta of mass greater than $10M_\odot$, 
$\tau$ is less than unity about
100 days after the explosion.
Since the rise time of Type II supernova is quite long, we think that if the 
time of core collapse is more than 100 days earlier than that of the phase
transition,
the $\gamma$-ray burst resulting from the fireball shock can then be detected.
This may apply to GRB980326 and GRB970228, as the Types of SNe 
associated with them are unknown.
Since in this case, the corresponding 
distance that the ejecta has reached is not large ($\sim1-3\times10^{16}$cm),
the fireball shock will definitely run into the dense ejecta not long 
after the burst and transit to 
a non-relativistic expansion regime (M${\acute{e}}sz{\acute{a}}$ros et al.
1998;
Dai $\&$ Lu 1999a,b), 
leading to a steeply decaying 
or even non-detectable afterglow, very similar to the ``SupraNova" model
(Vietri $\&$
Stella 1998) of 
$\gamma$-ray bursts. This agrees with the observations of GRB980326,
whose optical afterglow decays as $t^{-2.1}$ (Bloom et al. 1999).
Recent analysis (Galama et al. 1999) of the optical and near-infrared
afterglow of GRB970228 also shows
a steep temporal decay ($F_{\nu}\propto{t^{-1.73}}$).
On the other hand, the rise time of Type Ib/c supernova is much shorter
($\sim$2---3 weeks).
So, even at the time that the luminosity of the supernova begins to decline,
the scattering optical depth is still much greater than unity and no
significant amount
of gamma-rays can escape, unless there is a hole in the ejecta, which in
fact implies
a high-velocity moving jet in that direction resulted from 
a highly asymmetric supernova explosion.
Since a Type Ib/c supernova
SN1998bw has already been identified to be associated with GRB980425,
next we will discuss  in more detail this scenario.

In their models to explain the SN/GRB connection, Wang $\&$ Wheeler (1998),
Cen (1998) and Nakamura (1998) assumed that the highly asymmetric explosion
of Type Ib/c 
supernova makes the 
material in a small cone of the supernova ejecta  be preferentially first
blown out
of the deep gravitational potential well of the star. Immediately (about a few
seconds after the explosion), a tightly collimated jet from
the core collapse rushes through the preferred ``hole" and becomes an
ultra-relativistic jet after an expansion phase. However, because the
preexpelled
material may still run in the direction of the small cone, we speculate that
the fireball jet formed {\em immediately} after the explosion should be
slowed down
(viz. $\Gamma_0\ll100$) by the material before the fireball itself becomes
optically thin, and difficult to produce a $\gamma$-ray burst. (Anyway, the
mass in the
small cone is about $10^{-3}M_\odot$ even for
$\frac{\Omega}{4\pi}\sim10^{-3}$.)
But we think that this process may accelerate the preexpelled material in
the small cone  to 
mildly relativistic velocity ($\sim{0.8c}$) and leaves 
a preferred exit 
for the fireball formed {\em a few days later} by the conversion of a
newborn neutron star to
strange star or from the differentially rotating strange star. 
{\em In this case, the fireball jet can reach a large radius 
and produce a $\gamma$-ray burst
before catching up with the preexpelled material.} Moreover, at this time,
the preexpelled material will not scatter the gamma-ray photons
significantly, because
the scattering optical depth has decreased below unity.
In fact, there are some observational and theoretical evidences favouring
the existence of 
mildly relativistic jet in the supernova explosion:
1)The mysterious spot in SN1987A which appeared $5-7$ weeks after the
explosion at $\sim0.06$ arc-second away from the center, implies
a relativistic velocity (Rees 1987; Piran $\&$ Nakamura 1987) 
of $v\simeq(0.6\pm0.15)c/{\sin\alpha}$, where $\alpha$
is the angle between the velocity and the line of sight.
The new
analysis of SN1987A data provided stronger evidence for the original
``mystery spot" and  in addition a second spot on the opposite side of the
supernova,
suggesting relativistic jets (Nisenson $\&$ Papaliolios 1999). 
2) General relativistic numerical simulations (Piran $\&$ Nakamura 1987)
have demonstrated 
that collapse of the rapidly rotating core bounces along the rotation axis
to form jets moving with mildly relativistic velocity; 3)Superstrong magnetic
field formed immediately after the core collapse is claimed to
be able to punch a hole in the 
supernova ejecta and the preexpelled jet can also reach a relativistic 
velocity (Nakamura 1998).
Moreover, because the newborn
rapidly rotating neutron star may have a strong magnetic field, the energy
released through the magnetic dipole radiation can be as large as 
$\dot{E}\sim{B^{2}R^{6}\omega^{4}}/{c^{3}}\sim4\times10^{46}
{\rm ergs~s}^{-1}({B}/{10^{13}{\rm G}})^2({R}/{10^{6}{\rm cm}})^6({\omega}/
{10^{4}{\rm s}^{-1}})^4.$
Since a large fraction of this energy
will be converted to photons, the ensuing luminosity in fact exceeds the
Eddington
luminosity for a neutron star by about 8 orders of magnitude. This high
energy flow
and the inherent rotation of the ejecta(if the jet moves along the rotation
axis)
may maintain the emptiness of the ``hole".

\section{ Time scale and afterglows of GRBs associated with SNe}

It is  very likly that the ultra-relativistic radiation-dominated fireball, 
produced by the phase transition or/and
from the differentially rotating strange stars,
 becomes beamed to some extent by the MHD effect due to the anisotropy of the 
  configuration of the magnetic field  surrounding the strange stars.
As an estimate, we assume the enhancement factor is about 10 and thus the 
equivalent isotropic
energy $E_{i}$ of the fireball is of the order of $10^{53}$erg or even more.

How much of this fireball energy can flow out through the ``hole" is also
difficult to be determined. It may depend on the size of the ``hole" and the 
jet dynamics. But from the inferred total energy associated with GRBs, 
 which ranges from $10^{50}$erg to $10^{51.5}$erg (Freedman \& Waxman 1999),
 we estimate that only less than ten per cent of the fireball energy flowed
out
 the supernova ejecta and 
  produced the observed gamma-ray burst.
  For the case that 
only the phase transition
occurs, the $\gamma$-ray burst is more likely to be produced
by the external shocks 
 (Dermer 2000 and references therein) 
rather than internal shocks because the conversion of neutron
stars to strange stars is very quick and the energy deposition is impulsive.
The jet-like outflow expands outward in a 
way similar to a homogeneous fireball, sweeping up more and more external
matter.
We expect the existence of a stellar wind enviornment surrounding the massive
star GRB progenitor (Chevalier \& Li 1999).
External shocks will occur when the observer-frame energy of the swept-up
external
matter equals the initial energy 
of the fireball jet at a radius (M${\acute{e}}sz{\acute{a}}$ros 1999) 
\begin{equation}
r_{dec}\sim2\times10^{15}~{\rm cm}~{E_{i,53}}{\Gamma_{0,2}^{-2}}
(\dot{M}_{-5}/v_{w,3})^{-1},
\end{equation}
where $E_i=10^{53}E_{i,53}{\rm erg}$, $\Gamma_0=10^2\Gamma_{0,2}$,  
$\dot{M}=10^{-5}\dot{M}_{-5}{M_\odot}~{\rm yr^{-1}}$ and
$v_w=10^{3}v_{w,3}{\rm km~s^{-1}}$.
Here, $\dot{M}$ and $v_w$ are the mass-loss rate 
and velocity of the wind, respectively.
The duration of the GRBs in the observer's frame is
\begin{equation}
\Delta{t}\sim\frac{r_{dec}}{\Gamma_0^{2}c}\sim 10~{\rm s}~
E_{i,53}\Gamma_2^{-4}(\dot{M}_{-5}/v_{w,3})^{-1}.
\end{equation}
This time scale is consistent with the durations of those bursts  
 that are thought to connect with supernovae. 
Variability on time scales shorter than $\Delta{t}$ may occur on the cooling
time scale of electrons or on the dynamic scale for inhomogenities in the
external 
medium, but generally this is not ideal for reproducing highly variable 
profiles. Therefore, in this case we will generally see
bursts with simple profiles, agreeing well with 
GRB980425 and GRB980326 (Soffitta et al. 1998; Celidonio et al. 1998).
On the other hand, for GRB970228 that have a relatively complex time
structure, 
we think it may be produced by
internal shock resulting from  the differentially
rotation process of the newborn strange star, in which the faster shells catch
up and collide with the slower ones. 

The afterglows are generally expected to decay rapidly as the result of the
wind-shaped
circumburst enviornment (Chevalier \& Li 1999), which is consistent with the
observations of the GRB980329 and GRB970228. For the case of GRB980425,
it is  natural to think that  the jet geometry (i.e. we observe this jet
from the lateral
direction) makes us detect a weak gamma-ray intensity and
Its weak optical afterglow emission  is supposed to be
suppressed by the luminous optical radiation of SN1998bw and therefore not
seen
by us.
However when the radio afterglow dominates (we suggest the radio emission
is from
the afterglow of GRB980425 rather than SN1998bw), its emission angle
$\theta\sim\frac{1}{\gamma}\sim\frac{1}{2}$ is 
quite large (Wang et al. 1999) and at this time the observer may be inside
this angle; 
hence, the radio emission we received should be bright, considering
the short distance of the source from us.
 
Apart from the fraction of the fireball energy that flows out along the
directions of the high-velocity SN jets, 
the remaining, large portion of the  ultra-relativistic shell(s) will catch
up  and 
collide with the ejecta of the supernova that moves with a lower velocity
and be
immediately decelerated to a non-relativistic speed, heating the ejecta and
producing
super-Mev gamma-rays
at the same time. However, no significant amount of gamma-rays can escape
due to 
a high  scattering optical depth of the dense ejecta, i.e. 
$\tau=\sigma_{T}nl>1000$ (for the ejecta with $\sim5M_\odot$) at the time
two days after the explosion.
Therefore, a large fraction of the energy of the fireball shell 
will turn into the expansion energy
of the massive ejecta, leading to a supernova with very bright optical
luminosity and broad line emission, which are the very characteristics of
SN1998bw.

\section{Discussions}

We have suggested a possible explanation of the puzzling SN/GRB connection,
based on
the conversion of a newborn, massive neutron star to a strange star few days
after the supernova explosion. 
This suggestion is based on  a basic hypothesis that strange matter is the
true ground
state, which is entirely possible but not conclusive at present.
Thus we should only 
regard strange stars
as  possible stellar objects.
We think that the ultra-relativistic shell(s)
responsible for GRBs could not be produced by the supernova itself, but
possibly 
by the phase transition process or the differentially rotation process of the
newborn strange star.
The formation of a strange star requires that 
the total mass of the preconversion
neutron star should exceed $\sim1.8M_\odot$ for slowly rotating ones. 
According to the numerical simulation (Woosley et al. 1999), when supernova
occurs,
some matter may fail to  escape and fall back onto the neutron star.
For example, as they show, if the kinetic energy at infinity is set to be
about $1.2\times10^{51}$erg 
for a 25$M_\odot$ presupernova star,  about
$0.48M_\odot$ matter falls back onto the neutron star in about 1000 seconds.
Therefore, it is reasonable to believe that some supernova explosions,
especially
for the progenitors with moderately  massive mass ( perhaps in the range
$20-30M_\odot$),
could produce massive neutron stars. But, for very massive progenitors (
perhaps with
mass higher than $30M_\odot$), it is very likely that more than $1M_\odot$
matter falls
back and then the massive neutron star will promptly collapse to a black
hole, 
which then points to the
``collapsar" model or the two-step model (Cheng $\&$ Dai 1999) of GRBs
associated with
supernovae.

Our scenario clearly differs from previous discussions about the effects of
strange
quark matter on SN explosion (e.g. Benvenuto \& Horvath 1989; Dai, Peng \&
Lu 1995;
Gentile et al. 1993). In their works, the authors assumed a lower
deconfinement
density ($\sim~2-3\rho_0$) for soft EOSs and  strange stars were born
during the
 core collapse phase of the SN progenitors. The energy released 
 during the conversion may be a help to the SN
 explosion because of the increase of the shock-wave energy. 
But, we think that  this case is not appropriate for
 the production of GRBs due to
 the  baryon contamination of the massive envelope surrounding the strange
star. 
In our scenario, we suggested a  model involving the delayed birth of
strange stars,
in which the baryon contamination in this case is smaller and 
the mass of the newborn neutron stars can reach the critic value by accretion
from the supernova fallback matter. Thus, 
 we speculate that the SN explosion is normal while the associated GRBs are
powered
by the phase transition  which occurs only when the neutron stars
have spun down sufficiently. 

After all, note that this scenario has some distinct features or predictions.
First, a strange star, rather than a black hole or neutron star, is left 
after a SN/GRB event. 
 Since all non-strange matter is ejected during the burst, a bare strange star
is left. Thus in our galaxy, there may exist bare strange stars left by
past GRBs.
Discovery of this kind of interesting stellar objects will be a good
indication of our
proposed model. Certainly, if at later time the bare strange stars accreted
matter from 
outside, it may again be ``dressed" in a shell of ordinary matter. 
Second, 
 as the major part of the fireball shell collides with the supernova ejecta, 
a large fraction of the energy  released in the 
conversion will finally turn into the kinetic
energy of the ejecta, therefore the supernova should be very bright and
show broad emission lines. Thirdly, the afterglows of supernova-related
$\gamma$-ray bursts in this scenario will generally decay faster than usual
ones as the
result of  dense medium effect (Type II SNe) or interaction with the
surrounding
stellar wind (Type Ib/c SNe).
Finally,  prior to the phase transition, the  spin-down of the hot, newborn
neutron stars
 driven by the r-mode instability
 involves emission of up to $8\times10^{52}$erg of gravitational waves, making
 the gravitational radiation potentially observable.

\acknowledgements 
 We would like to thank the referee for his valuble suggestions.
This work was supported by the National Natural Science Foundation 
and the Foundation of the Ministry of Education of China.

\end{document}